\documentclass[10pt,conference]{IEEEtran}
\usepackage{amssymb,amsmath,epsfig,graphicx,theorem,threeparttable}
\usepackage{amsfonts}
\usepackage{bm}
\usepackage{cite}

\newtheorem{Theo}{Theorem}
\newtheorem{Lem}{Lemma}

\begin{document}
\title{Optimal Distortion-Power Tradeoffs in Gaussian Sensor Networks
\thanks{This work was supported by NSF Grants CCR $03$-$11311$, CCF $04$-$47613$ and CCF $05$-$14846$;
and ARL/CTA Grant DAAD $19$-$01$-$2$-$0011$.}}

\IEEEoverridecommandlockouts

\author{Nan Liu \qquad Sennur Ulukus \\
\normalsize Department of Electrical and Computer Engineering \\
\normalsize University of Maryland, College Park, MD 20742 \\
\normalsize {\it nkancy@umd.edu} \qquad {\it ulukus@umd.edu} }

\maketitle

\begin{abstract}
We investigate the optimal performance of dense sensor networks by
studying the joint source-channel coding problem. The overall goal
of the sensor network is to take measurements from an underlying
random process, code and transmit those measurement samples to a
collector node in a cooperative multiple access channel with
imperfect feedback, and reconstruct the entire random process at
the collector node. We provide lower and upper bounds for the
minimum achievable expected distortion when the underlying random
process is Gaussian. In the case where the random process
satisfies some general conditions, we evaluate the lower and upper
bounds explicitly and show that they are of the same order for a
wide range of sum power constraints. Thus, for these random
processes, under these sum power constraints, we determine the
achievability scheme that is order-optimal, and express the
minimum achievable expected distortion as a function of the sum
power constraint.
\end{abstract}

\section{Introduction}
With the recent advances in the hardware technology, small cheap
nodes with sensing, computing and communication capabilities have
become available. In practical applications, it is possible to
deploy a large number of these nodes to sense the environment.
In this paper, we investigate the optimal performance of a dense
sensor network by studying the joint source-channel coding
problem. The sensor network is composed of $N$ sensors, where $N$
is very large, and a single collector node. The overall goal of
the sensor network is to take measurements from an underlying
random process $S(t)$, $0 \leq t \leq T_0$, code and transmit
those measured samples to a collector node in a cooperative
multiple access channel with imperfect feedback, and reconstruct
the entire random process at the collector node. We investigate
the minimum achievable expected
 distortion and the corresponding achievability scheme when the underlying random process is
Gaussian and the communication channel is a cooperative Gaussian
multiple access channel with feedback (potentially imperfect).

Scaglione and Servetto \cite{Servetto:2002} investigated the
scalability of the sensor networks. The goal of the sensor network
in \cite{Servetto:2002} was that each sensor reconstructs the data
measured by all of the sensors using sensor broadcasting. In this
paper, we focus on the case where the reconstruction is required
only at the collector node. Also, in this paper, the task is not
the reconstruction of the data the sensors measured, but the
reconstruction of the underlying random process.

Gastpar and Vetterli \cite{Gastpar:sensor2005} studied the case
where the sensors observe a noisy version of a linear combination
of $L$ Gaussian random variables which all have the same variance,
code and transmit those observations to a collector node, and the
collector node reconstructs the $L$ random variables. In
\cite{Gastpar:sensor2005}, the expected distortion achieved by
applying separation-based approaches was shown to be exponentially
worse than the lower bound on the minimum achievable expected
distortion. In this paper, we study the case where the data of
interest at the collector node is not a finite number of random
variables, but a random process. We assume that the sensors are
able to take noiseless samples.
 Our upper bound is also developed by
using a separation-based approach, but it is shown to be of the
same order as the lower bound, for a wide range of power
constraints for random processes that satisfy some general
conditions.

El Gamal \cite{ElGamal:2005} studied the capacity of dense sensor
networks and found that all spatially band-limited Gaussian
processes can be estimated at the collector node, subject to any
non-zero constraint on the mean squared distortion. In this paper,
we study the minimum achievable expected distortion for
space-limited, and thus, not band-limited, random processes, and
we show that the minimum achievable expected distortion decreases
to zero as the number of nodes increases, unless the sum power
constraint is unusually small.

In \cite{Liu_Ulukus:2006}, we studied this problem when the
underlying process was Gauss-Markov. We found the minimum
achievable expected distortion and the corresponding order-optimal
achievability scheme for
 a wide range of sum power constraints.
In this work, we extend our results to the case where the
underlying process is a Gaussian random process which satisfies
some general conditions. We first provide lower and upper bounds
for the minimum achievable expected distortion for arbitrary
Gaussian random processes for which the Karhunen-Loeve expansion
exists. Then, we focus on the case where the Gaussian random
process also satisfies some general conditions, evaluate the lower
and upper bounds explicitly, and show that they are of the same
order, for a wide range of sum power constraints. Thus, for these
random processes, under a wide range of sum power constraints, we
determine an order-optimal achievability scheme, and identify the
minimum achievable expected distortion.

\section{System Model} \label{systemmodel}
The collector node wishes to reconstruct a sequence of random
processes $\{S^{(l)}(t), 0 \leq t \leq T_0\}_{l=1}^ n$, i.i.d. in
time, where $l$ denotes the time, $n$ is the block length, and $t$
denotes the spatial position. At each time instant, the random
process $S(t)$ is assumed to be Gaussian with zero-mean and a
continuous autocorrelation function $K(t,s)$. The $N$ sensor nodes
are placed at positions $0=t_1 \leq t_2 \leq \cdots \leq t_N=T_0$,
and observe samples $\mathbf{S}_N=(S(t_1),S(t_2),\cdots,S(t_N))$.
For simplicity and to avoid irregular cases, we assume that the
sensors are equally spaced.
The distortion measure is the squared error,
\begin{align}
d(s(t),\hat{s}(t)) = \frac{1}{T_0}\int_{0}^{T_0}
(s(t)-\hat{s}(t))^2 dt
\end{align}

Each sensor node and the collector node, denoted as node 0, is
equipped with one transmit and one receive antenna. At any time
instant, let $X_i$ and $Y_i$ denote the signals transmitted by and
received at, node $i$, and let $h_{ji}$ denote the channel gain
from node $j$ to node $i$. Then, the received signal at node $i$
can be written as,
\begin{align}
Y_i=\sum_{j=0,j \neq i}^N h_{ji} X_j+Z_i, \qquad i=0,1,2,\cdots,N
\end{align}
where $\{Z_i\}_{i=0}^N$ is a vector of $N+1$ independent and
identically distributed, zero-mean, unit-variance Gaussian random
variables. Therefore, the channel model of the network is such
that all nodes hear a linear combination of the signals
transmitted by all other nodes at that time instant. We assume
that the channel gain $h_{ij}$ is bounded, i.e.,
\begin{align}
\bar{h}_l \leq h_{ij} \leq \bar{h}_u, \qquad i,j=0,1,\cdots,N
\end{align}
where $\bar{h}_u$ and $\bar{h}_l$ are positive constants
independent of $N$. This model is quite general and should be
satisfied easily: by the conservation of energy, $h_{ij}^2 \leq 1,
i,j=0,1,\cdots,N$, and since all nodes are within finite distance
from each other, the channel gains should be lower bounded as
well. This channel gain model differs from the one in
\cite{Liu_Ulukus:2006}, and is more reasonable. Since the
collector node is also equipped with a transmit antenna, there is
feedback from the collector node to the sensor nodes. However, in
a wireless environment, this feedback is imperfect, in the sense
that it is corrupted by channel noise, as well as interference
from other simultaneously transmitting nodes.

We assume that $K$ channel uses are allowed per realization of the
random process in time for the reconstruction, where $K$ is a
finite positive integer independent of the number of sensors $N$
and block length $n$. We also assume that all sensors share a sum
power constraint of $P(N)$ which is a function of $N$.
The two most interesting cases for the sum power constraint are
$P(N)=N P_{\text{ind}}$ where each sensor has its individual power
constraint $P_{\text{ind}}$, and $P(N)=P_{\text{tot}}$ where all
sensors share a constant total power constraint $P_{\text{tot}}$.
Our goal is to determine the scheme that achieves the minimum
achievable expected distortion $D^N$ at the collector node for a
given total transmit power constraint $P(N)$, and also to
determine the rate at which this distortion goes to zero as a
function of the number of sensor nodes and the power constraint.

All logarithms are base $e$. Due to space limitations, all proofs
are omitted here and can be found in \cite{Liu_Ulukus:2005}.

\section{The Class of Gaussian Random Processes}\label{defineA}

For a Gaussian random process $S(t)$ with continuous
autocorrelation, we perform the Karhunen-Loeve expansion
\begin{align}
S(t)=\sum_{k=0}^\infty S_k \phi_k(t)
\end{align}
 to obtain the
ordered eigenvalues $\{\lambda_k\}_{k=0}^\infty$, and
corresponding eigenfunctions $\{\phi_k(t), t \in
[0,T_0]\}_{k=0}^\infty$.

Let $\mathcal{A}$ be the set of Gaussian random processes that
satisfy the following conditions:

1. There exist nonnegative constants $d_l$, $d_u$ and nonnegative
integers $c_l$, $c_u$ and $K_0 \geq c_u+1$ such that when $k >
K_0$,
\begin{align}
\frac{d_l}{(k+c_l)^x} \leq \lambda_k \leq \frac{d_u}{(k-c_u)^x}
\label{decrate}
\end{align}
for some constant $x>1$. The condition that $x>1$ is without loss
of generality, because for all continuous autocorrelations, the
eigenvalues decrease faster than $k^{-1}$.

2. In addition to continuity, $K(t,s)$ satisfies the Lipschitz
condition of order $1/2 < \alpha \leq 1$, i.e., there exists a
constant $B$ such that
\begin{align}
|K(t_1,s_1)-K(t_2,s_2)| \leq B
\left(\sqrt{(t_1-t_2)^2+(s_1-s_2)^2}\right)^\alpha
\end{align}
for all $t_1,s_1,t_2,s_2 \in [0,T_0]$.

3. For $k=0,1,\cdots$, the function $\phi^2_k(s)$ and the function
$K(t,s)\phi_k(s)$ as a function of $s$ satisfy the following
condition: there exist positive constants $B_1$, $B_2$, $B_3$, $B_4$,
$\beta \leq 1$, $\gamma \leq 1$, and nonnegative constant $\tau$, independent of $k$ such that
\begin{align}
|\phi^2_k(s_1)-\phi^2_k(s_2)|  \leq B_3(k+B_4)^\tau
|s_1-s_2|^\gamma
\end{align}
and for all $t \in [0,T_0]$
\begin{align}
|K(t,s_1)\phi_k(s_1)-K(t,s_2)\phi_k(s_2)|  \leq B_2(k+B_1)^\tau
|s_1-s_2|^\beta
\end{align}
for all $s_1,s_2 \in [0,T_0]$.


The reasons why these conditions are needed for explicit
evaluation of the lower and upper bounds on the minimum achievable
expected distortion
 will be clear from the proofs, which are not presented here due to space limitations.
We provide some intuition as to why these conditions are needed.
Condition 1 shows that we consider random processes that have
eigenvalues $\lambda_k$ which decrease at a rate of approximately
$k^{-x}$. The lower (upper) bound on the eigenvalues in
(\ref{decrate}) will be used to calculate the lower (upper) bound
on the minimum achievable expected distortion. Conditions 2 and 3
are needed because instead of the random process itself that is of
interest to the collector node, the collector node, at best, can
know only the sampled values of the random process. How well the
entire process can be approximated from its samples is critical in
obtaining quantitative results. Lipschitz conditions describe the
quality of this approximation well. For instance, by condition 3,
we require the variation in the eigenfunctions $\phi_k^2$ to be no
faster than order $k^\tau$. We note that the well-known
trigonometric basis satisfies this condition. We also note that
our conditions are quite general. Many random processes satisfy
these conditions including the Gauss-Markov process, Brownian
motion process, etc.

\section{A Lower Bound on the Minimum Achievable Expected Distortion} \label{seclowerbound}
\subsection{Arbitrary Gaussian Random Processes}
The lower bound is obtained by assuming
 that all of the sensor nodes know the random process exactly, and,
the sensor network forms an $N$-transmit 1-receive antenna
point-to-point system to transmit the random process to the
collector node. Let $C_u^N$ be the capacity of this point-to-point
system and $D_p(R)$ be the distortion-rate function of the random
process $S(t)$ \cite{Berger:book}. In this point-to-point system,
the
 separation principle holds and feedback, imperfect or perfect, does not increase the capacity, and therefore
\begin{align}
D^N \geq D_p(C_u^N)
\end{align}
To evaluate $D_p(C_u^N)$, we first find the rate distortion
function, which is the inverse function of $D_p(R)$, of $S(t)$
\cite[Section 4.5]{Berger:book} as,
\begin{align}
R(\theta)= \sum_{k=0}^\infty \max \left(0, \frac{1}{2} \log
\left(\frac{\lambda_k}{\theta} \right) \right) \label{rate}
\end{align}
and
\begin{align}
D(\theta)=T_0^{-1} \sum_{k=0}^\infty \min (\theta,
\lambda_k)\label{distortion}
\end{align}
It can be seen that the function $R(\theta)$ is a strictly
decreasing function of $\theta$ when $\theta \leq \lambda_0$.
Hence, in this region, the inverse function of $R(\theta)$ exists,
which we will call $\theta(R), R \geq 0$. Next, we find $C_u^N$,
the capacity of the $N$-transmit 1-receive antenna point-to-point
system
 \cite{Telatar:1999} over $K$ channel uses as,
\begin{align}
C_u^N=\frac{K}{2} \log \left(1+P(N)\sum_{i=1}^N h^2 _{i0}  \right)
\label{Cupper}
\end{align}
Then, we have
\begin{align}
D_p(C_u^N)=D(\theta(C_u^N)) \label{eqn}
\end{align}
Thus, for arbitrary Gaussian random processes, a lower bound on
the minimum achievable expected distortion is
\begin{align}
D_l^N=D_p(C_u^N) \label{ulukus1}
\end{align}

\subsection{The Class of Gaussian Random Processes in $\mathcal{A}$}

Next, we evaluate $D_p(C_u^N)$ for the class of Gaussian random
processes in $\mathcal{A}$. We will divide our discussion into two
separate cases based on the sum power constraint. For the first
case, $P(N)$ is such that
\begin{align}
\lim_{N \rightarrow \infty} \left(NP(N)\right)^{-1}=0
\label{powerconstraint}
\end{align}
is satisfied. The cases where $P(N)=N P_{\text{ind}}$ and
$P(N)=P_{\text{tot}}$ are included in $P(N)$ satisfying
(\ref{powerconstraint}). Note that $C_u^N$ increases monotonically
in $N$. Since we are interested in the number $\theta(C_u^N)$, we
will analyze the function $\theta(R)$ where $R$ is very large.
Correspondingly, this means that we will analyze the function
$D(\theta)$ where $\theta$ is very small.


\begin{Lem} \label{cut1}
For large enough $R$, we have
\begin{align}
\theta(R) \geq d_l \left(\frac{x}{4} \right)^x R^{-x}
\label{newstar}
\end{align}
\end{Lem}

\begin{Lem} \label{cut5}
For small enough $\theta$, we have
\begin{align}
D(\theta) & \geq \frac{d_l^{\frac{1}{x}}}{2T_0}
\theta^{1-\frac{1}{x}} \label{largeN2}
\end{align}
\end{Lem}

We are now ready to calculate the distortion. When $N$ is large
enough and the sum power constraint $P(N)$ satisfies
(\ref{powerconstraint}), using (\ref{largeN2}), (\ref{newstar})
and the fact that
\begin{align}
C_u^N \leq \frac{K}{2} \log \left(1+\bar{h}_u^2 NP(N) \right)
\end{align}
%
a lower bound on the achievable distortion is
\begin{align}
\Theta \left(\left(\log(NP(N))\right)^{1-x} \right)
\label{ulukus123}
\end{align}

For the second case, $P(N)$  is such that (\ref{powerconstraint})
is not satisfied.
 $C_u^N$ is either a constant independent of
 $N$ or goes to zero as $N$ goes to infinity. Examining (\ref{rate}), we see that $\theta(C_u^N)$ is bounded
 below by a constant independent of $N$, and hence, $D\left(\theta\left(C_u^N\right)\right)$ is a constant and does not
 go to zero as $N$ increases. Therefore, combining this case with (\ref{ulukus123}), for all possible power constraints $P(N)$, a lower
bound on the distortion is
\begin{align}
\Theta \left(  \min \left(\left(\log(NP(N)) \right)^{1-x}, 1
\right)  \right)
\end{align}

When the sum power constraint grows almost exponentially with the
number of nodes, the lower bound on the minimum achievable
expected distortion decreases inverse polynomially with $N$. Even
though this provides excellent performance, it is impractical
since sensor nodes are low energy devices and it is often
difficult, if not impossible, to replenish their batteries.

When the sum power constraint is such that (\ref{powerconstraint})
is not satisfied, the transmission power is so low that the
communication channels between the sensors and the collector node
are as if they do not exist. The estimation error is on the order
of 1, which is equivalent to the collector node blindly estimating
$S(t)=0$ for all $t \in [0, T_0]$. Even though the consumed power
$P(N)$ is very low in this case, the performance of the sensor
network is unacceptable; even the lower bound on the minimum
achievable expected distortion does not decrease to zero with the
increasing number of nodes.

For practically meaningful sum power constraints, including the
cases of $P(N)=N P_{\text{ind}}$ and $P(N)=P_{\text{tot}}$, the
lower bound on the minimum achievable expected distortion decays
to zero at the rate of
\begin{align}
\left(\log N\right)^{1-x} \label{ulukus7}
\end{align}

\section{An Upper Bound on the Minimum Achievable Expected Distortion} \label{secupperbound}
\subsection{Arbitrary Gaussian Random Processes}
Any distortion found by using any achievability scheme will serve
as an upper bound on the minimum achievable expected distortion.
We consider the following separation-based achievable scheme:
First, we perform distributed rate-distortion coding at all sensor
nodes using \cite[Theorem 1]{Flynn:1987}. After obtaining the
indices of the rate-distortion codes, we transmit the indices as
independent messages using the amplify-and-forward method
introduced in \cite{Gastpar:2005}. The distortion obtained using
this scheme will be denoted as $D_u^N$.

We apply \cite[Theorem 1]{Flynn:1987}, generalized to $N$ sensor
nodes in \cite[Theorem 1]{Chen:2004}, to obtain an achievable
rate-distortion point.
\begin{Theo}
If the individual rates are equal, the following sum rate and
distortion are achievable,
\begin{align}
& D_a^N(\theta') \nonumber \\
&  = \frac{1}{T_0}\int_{0}^{T_0}
\left(K(t,t)-\frac{T_0}{N-1}\bm{\rho}_N^T(t)
\left(\Sigma_N'+\theta'
I\right)^{-1} \bm{\rho}_N(t) \right) dt \\
& R_a^N(\theta')  =\sum_{k=0}^{N-1} \frac{1}{2} \log
\left(1+\frac{\mu_k^{(N)'}}{\theta'} \right)
\end{align}
where $\bm{\rho}_N(t)$ is an $N \times 1$ column vector with the
$i$-th entry being $K\left(t, (i-1)/(N-1)T_0 \right)$. $\Sigma_N'$
is an $N \times N$ matrix with the $(i,j)$-th entry being
$T_0/(N-1)$ $K \left((i-1)/(N-1) T_0, (j-1)/(N-1) T_0 \right)$,
and $\mu_0^{(N)'},\mu_1^{(N)'},\cdots,\mu_{N-1}^{(N)'}$ are the
eigenvalues of $\Sigma_N'$.
\end{Theo}
We further evaluate $D_a^N(\theta')$ in the next lemma.
\begin{Lem} \label{whatnoname}
For all Gaussian random processes, we have
\begin{align}
&D_a^N(\theta') \nonumber \\
& = O \left(\max \left(A^{(N)}, B^{(N)}, \frac{1}{T_0}
\sum_{k=0}^{N-1} \left(\frac{1}{\theta'}+\frac{1}{\mu_k^{(N)'}}
\right)^{-1} \right) \right)
\end{align}
where $A^{(N)}$ and $B^{(N)}$ are defined as
\begin{align}
&A^{(N)} = \frac{2}{T_0}\sum_{i=1}^{N-1}
\int_{\frac{i-1}{N-1}T_0}^{\frac{i}{N-1}T_0}
\left(\bm{\rho}_N\left(\frac{i-1}{N-1}T_0 \right)-\bm{\rho}_N(t) \right)_i dt \nonumber \\
& +\frac{1}{T_0}\sum_{i=1}^{N-1}
\int_{\frac{i-1}{N-1}T_0}^{\frac{i}{N-1}T_0} \left(K(t,t)-K
\left(\frac{i-1}{N-1}T_0,\frac{i-1}{N-1}T_0\right) \right) dt
\end{align}
and
\begin{align}
B^{(N)} =\frac{2}{T_0}\sum_{i=1}^{N-1}
\int_{\frac{i-1}{N-1}T_0}^{\frac{i}{N-1}T_0}
\left|\left|\bm{\rho}_N\left(\frac{i-1}{N-1}T_0
\right)-\bm{\rho}_N(t) \right|\right| dt
\end{align}
respectively.
\end{Lem}
Lemma \ref{whatnoname} tells us that the expected distortion
achieved by using the separation-based scheme is upper bounded by
the maximum of three types of distortion. We define the third
distortion as
\begin{align}
D_b^N(\theta') = \frac{1}{T_0} \sum_{k=0}^{N-1}
\left(\frac{1}{\theta'}+\frac{1}{\mu_k^{(N)'}} \right)^{-1}
\end{align}

Now, we determine an achievable rate for the communication channel
from the sensor nodes to the collector node. The channel in its
nature is a multiple access channel with potential cooperation
between the transmitters and imperfect feedback  from the
collector node. The capacity region for this channel is not known.
We get an achievable sum rate, with identical individual rates,
for this channel by using the idea presented in
\cite{Gastpar:2005}. This result is a generalization of
\cite[Theorem 1]{ElGamal:2005} from a constant power constraint to
a more general power constraint.
\begin{Theo} \label{generalelgamal}
When the sum power constraint $P(N)$ is such that there exists an
$\epsilon > 0$ where
\begin{align}
\lim_{N \rightarrow \infty} P(N) N^{\frac{1}{2}-\epsilon}>1
\label{powersad1}
\end{align}
the following sum rate is achievable over $K$ channel uses,
\begin{align}
C_a^N=\frac{d_0 K}{2} \log (NP(N))
\end{align}
where $d_0$ is a positive constant, independent of $N$. Otherwise,
$C_a^N$ approaches a positive constant or zero as $N \rightarrow
\infty$.
\end{Theo}
Theorem \ref{generalelgamal} shows that when the sum power
constraint is such that (\ref{powersad1}) is satisfied, the
achievable rate increases with $N$. Otherwise, the achievable rate
is either a positive constant or decreases to zero, which will
result in poor estimation performance at the collector node.

The function $R_a^N\left(\theta'\right)$ is a strictly decreasing
function of $\theta'$, thus, the inverse function exists, which we
will denote as $\theta_a^N(R)$. Hence, to find $D_u^N$, we first
find $ \theta_a^N\left(C_a^N\right)$, and then,
\begin{align}
D_u^N & = D_a^N \left(\theta_a^N \left(C_a^N\right) \right)
\label{sigmaDeqn2}
\end{align}
We will perform this calculation when the underlying random
process is in $\mathcal{A}$.

\subsection{The Class of Gaussian Random Processes in $\mathcal{A}$}

We analyze all three types of distortion for Gaussian random
processes in $\mathcal{A}$.

\begin{Lem} \label{addgeneral}
For Gaussian random processes in $\mathcal{A}$, we have
\begin{align}
A^{(N)}&= O \left(N^{-\alpha} \right) \\
B^{(N)} & = O \left(N^{\frac{1}{2}-\alpha} \right)
\end{align}
\end{Lem}
The results in Lemma \ref{addgeneral} depend crucially on
condition 2 in the definition of $\mathcal{A}$ in Section
\ref{defineA}.

Using the results of Lemma \ref{addgeneral}, we have
\begin{align}
D_u^N= O \left(\max \left(N^{\frac{1}{2}-\alpha}, D_b^N \left(\theta_a^N \left(C_a^N\right) \right)
\right) \right) \label{markovvalid}
\end{align}
It remains to evaluate $D_b^N \left(\theta_a^N \left(C_a^N\right)
\right)$. We first define two sequences $\vartheta_L^N$ and
$\vartheta_U^N$ which satisfy
\begin{align}
\lim_{N \rightarrow \infty} \frac{1}{\vartheta_L^N N^{\min
\left(x, \frac{x \gamma}{\tau}, \frac{\alpha x}{x-1}, \frac{\beta
x}{x+\tau} \right)}} =0, \quad \lim_{N \rightarrow \infty}
\vartheta_U^N =0 \label{increasing}
\end{align}
\begin{Lem} \label{nothingleft}
For large enough $N$ and $R$ in the interval of
\begin{align}
\left[\frac{2 d_u^{\frac{1}{x}} x^2}{x-1}
 \left(\vartheta_U^N \right)^{-\frac{1}{x}}, \frac{x
d_l^{\frac{1}{x}}}{8} \left(\vartheta_L^N
\right)^{-\frac{1}{x}}\right]  \label{partialinterval}
\end{align}
we have
\begin{align}
\frac{x^x d_l}{8^x R^x} \leq \theta_a^N(R) \leq \frac{2^x x^{2x}
d_u}{(x-1)^x R^x}  \label{partial}
\end{align}
\end{Lem}
Hence, for all $P(N)$ that satisfy
\begin{align}
 \lim_{N \rightarrow \infty} \frac{NP(N)}{e^{N^{\min
 \left(1, \frac{ \gamma}{\tau}, \frac{\alpha }{x-1}, \frac{\beta }{x+\tau} \right)}}} =0 \label{onlycase}
\end{align}
and (\ref{powersad1}), we have $C_a^N$ in the interval of
(\ref{partialinterval}) and Lemma \ref{nothingleft} applies.

Now we upper bound $D_b^N(\theta')$.
\begin{Lem} \label{somethingleft}
For $\theta' \in [\vartheta_L^N, \vartheta_U^N]$ for large enough
$N$, we may upper bound $D_b^N(\theta')$ as
\begin{align}
D_b^N(\theta') \leq \frac{4 d_u^{\frac{1}{x}}}{T_0}
\left(\frac{x+1}{x-1}\right) \theta'^{1-\frac{1}{x}}
\label{nuoconclude1}
\end{align}
\end{Lem}
Proofs of Lemmas \ref{nothingleft} and \ref{somethingleft} use all
conditions 1, 2 and 3 in the definition of $\mathcal{A}$ in
Section \ref{defineA}.

Hence, when $P(N)$ is such that (\ref{onlycase}) and
(\ref{powersad1}) are satisfied, using (\ref{nuoconclude1}) and
(\ref{partial}) and the fact that when $R$ is in the interval of
(\ref{partialinterval}), $\theta_a^N(R)$ is in $[\vartheta_L^N,
\vartheta_U^N]$,
 we have
\begin{align}
& D_b^N (\theta_a^N (C_a^N))\leq \Theta \left( \left( \log (NP(N))
\right)^{1-x} \right)
\end{align}
From (\ref{markovvalid}), an upper bound on the minimum achievable
expected distortion is
\begin{align}
\Theta \left( \left( \log (NP(N)) \right)^{1-x} \right)
\label{upperboundrepeat}
\end{align}
when the sum power constraint satisfies (\ref{powersad1}) and
\begin{align}
\lim_{N \rightarrow \infty} \frac{NP(N)}{e^{N^{\min
 \left(1, \frac{ \gamma}{\tau}, \frac{\alpha }{x-1},
 \frac{2 \alpha-1}{2(x-1)}, \frac{\beta }{x+\tau} \right)}}}=0 \label{powerfinalhope}
\end{align}

This upper bound on the minimum achievable expected distortion
coincides with the the lower bound when $P(N)$ satisfies
(\ref{powersad1}) and (\ref{powerfinalhope}). The practically
interesting cases of $P(N)=N P_{\text{ind}}$ and
$P(N)=P_{\text{tot}}$ fall into this region. In both of these
cases, the minimum achievable expected distortion decreases to
zero at the rate of
\begin{align}
\left(\log N \right)^{1-x}
\end{align}
which is the same as the lower bound in (\ref{ulukus7}).

\section{Conclusion}
We investigate the performance of dense sensor networks by
studying the joint source-channel coding problem. We provide lower
and upper bounds for the minimum achievable expected distortion
when the underlying random process is Gaussian. When the random
process satisfies some general conditions, we evaluate the lower
and upper bounds explicitly, and show that they are both of order
$\left( \log (NP(N)) \right)^{1-x} $
 for a wide range of sum power constraints
ranging from $N^{-\frac{1}{2}}$ to $\frac{e^{N^{\min
 \left(1, \frac{ \gamma}{\tau}, \frac{\alpha }{x-1},
 \frac{2 \alpha-1}{2(x-1)}, \frac{\beta }{x+\tau} \right)}}}{N}$.
 Therefore, for random processes that satisfy these general
conditions, and under these sum power constraints, we have found
that an order-optimal scheme is a separation-based scheme, that is
composed of distributed rate-distortion source coding
\cite{Flynn:1987} and amplify-and-forward channel coding
\cite{Gastpar:2005}, and the imperfect feedback link does not need
to be utilized.

The most interesting cases of $P(N)=N P_{\text{ind}}$ and
$P(N)=P_{\text{tot}}$ fall into this region, and for these cases,
the minimum achievable expected distortion decreases to zero at
the rate of $\left( \log N\right)^{1-x}$. Hence, the power
constraint $P(N)=P_{\text{tot}}$ performs as well as $P(N)=N
P_{\text{ind}}$ ``order-wise'', and therefore, in practice we may
prefer to choose $P(N)=P_{\text{tot}}$. In fact, any $P(N) \geq
N^{-1/2+\epsilon}$ , for example,
$P(N)=N^{-1/3}$, performs as well as $P(N)=N P_{\text{ind}}$ and
$P(N)=P_{\text{tot}}$. This means that if we choose $P(N)=N^{-1/3}$ , as we add more and more sensors, the sum
power constraint can go to zero without affecting the performance
``order-wise''.

\bibliographystyle{unsrt}
\bibliography{ref}

\end{document}